\documentclass[fleqn, twoside]{article}
\usepackage{espcrc2}
\usepackage{graphicx}

\title {Using Approximating Polynomials in Partial-Global Dynamical 
Simulations}

\author{Andrei Alexandru\address[UOFC]{Department of Physics, 
University of Colorado, Boulder, CO 80309-0390} and Anna Hasenfratz
\addressmark[UOFC]}

\begin{document}

\begin{abstract}
Smeared link fermionic actions can be straightforwardly simulated with 
partial-global updating. The efficiency of this simulation is greatly 
increased if the fermionic matrix is written as a product of several 
near-identical terms. Such a break-up can be achieved using polynomial 
approximations for the fermionic matrix. In this paper we will focus on 
methods of determining the optimum polynomials.
\end{abstract}

\maketitle

\section{MOTIVATION}

Dynamical fermion simulations with smeared links are used more and more 
frequently. Smeared links improve the chiral and flavor symmetry of 
Wilson and staggered fermionic actions (see references in \cite{anna4}). 
The main obstacle in
using these actions is finding efficient algorithms to simulate them. In this 
paper we present some techniques that are used to improve the efficiency of 
these simulations. We will focus here on the HYP action \cite{anna4,anna1,anna2,anna3},
but we believe that these methods are more general and can be used for
different actions as well.

The HYP action \cite{anna2} is a modified staggered fermion action
\begin{equation}
S=S_{G}\left(U\right)+\bar{S}_{G}(V)+S_{F}\left(V\right)
\end{equation}
where $S_{G}$ is the traditional pure gauge action defined in terms of thin links $U$,
$\bar{S}_G$ is a gauge action defined in terms of smeared links $V$ and
\begin{equation}
S_{F}  =  -\frac{n_{f}}{4}\ln \det \Omega,
\end{equation}
where $\Omega = M^\dagger M$ restricted to even sites,
is the staggered fermion action defined in terms of HYP links $V$ rather than thin links.
The role of $\bar{S}_G$ is discussed in \cite{anna2}.

To simulate this action we use a two step approach. We first update a part of the thin links,
$U\rightarrow U'$, using a heath-bath and/or an over-relaxation step, then we smear
the thin links using the hyper-cubic (HYP) blocking \cite{anna1} 
and compute the fat link action, $S_G\left(V'\right)+S_F\left(V'\right)$.
We accept this new configuration with probability
\begin{equation}
P_{acc}=\min \left\{ 1,e^{-\bar{S}_{G}\left(V'\right)+\bar{S}_{G}\left(V\right)-
S_{F}\left(V'\right)+S_{F}\left(V\right)}\right\}.
\end{equation}

The most difficult step in the algorithm is computing the determinant ratio
\begin{eqnarray}
e^{-S_{F}\left(V'\right)+S_{F}\left(V\right)} = 
\left(\frac{\det \Omega \left(V'\right)}{\det \Omega \left(V\right)}\right)^{n_{f}/4} \nonumber\\
= \frac{\int d\xi d\xi ^{\dagger }e^{-\xi ^{\dagger }\left(\Omega^{\prime -1/2}\Omega 
\Omega ^{\prime -1/2}\right)^{n_{f}/4}\xi }}{\int d\xi d\xi ^{\dagger }e^{-\xi ^{\dagger }\xi }}.
\end{eqnarray}
Instead of computing the ratio 
itself we use a stochastic estimator suggested by the last part of the relation above
\begin{equation}
E_{A}\left[\xi \right]  =  e^{-\xi ^{\dagger }\left(A-1\right)\xi },
\end{equation}
where $A=\left(\Omega ^{\prime -1/2}\Omega \Omega ^{\prime -1/2}\right)^{n_{f}/4}$. 
This estimator averages to the determinant ratio since $A$ is hermitian and positive definite.
Furthermore, the variance of this estimator is
\begin{equation}
\sigma ^{2}\left(E_{A}\right) = \frac{1}{\det \left(2A-1\right)}-\frac{1}{\left(\det A\right)^{2}}.
\end{equation}
The relation above holds only if the spectrum of $A$ is bounded from below by 1/2 otherwise the variance 
diverges. This condition is not necessarily fulfilled by our matrix. To address this problem we modify
our stochastic estimator
\begin{equation}
E'_{A}\left[\xi \right]  =  e^{-\sum_{j=1}^n \xi_j ^{\dagger }\left(A^{1/n}-1\right)\xi_j }. \label{estimator}
\end{equation}
This estimator has a finite variance if the lowest eigenvalue of $A$ is greater than $1/2^n$. This 
condition can be fulfilled if we choose $n$ properly. However, this forces us to compute the $n^{th}$
root of fermionic matrices. This is why we are forced to use the polynomial approximation.

In order to make the algorithm more efficient we employed a reduced matrix \cite{anna1,anna2,anna3}:
$\Omega_r = \Omega e^{-2 f\left(\Omega\right)}$,
where $f$ is a cubic function. The fermionic part will be $S_F=\frac{n_f}{4} \ln \det \Omega_r$ and the 
reminder is absorbed in $\bar{S}_G$.

Lastly, we mention that it can be proved that the algorithm satisfies the detailed balance condition
for either of the estimators presented above.

\section{POLYNOMIAL APPROXIMATION} 

To compute the estimator (\ref{estimator}), where 
$A=\left(\Omega_r^{\prime-1/2} \Omega_r \Omega_r^{\prime-1/2}\right)^{n_f/4}$, 
we need to find polynomial approximations for the matrix functions
involved in the estimator
\begin{eqnarray}
P^{\left(n\right)}_k\left(\Omega\right) &\simeq & F\left(\Omega\right)^{-1/2n}, \nonumber\\
Q^{\left(n\right)}_k\left(\Omega\right) &\simeq & F\left(\Omega\right)^{1/n},
\end{eqnarray}
where $F\left(\Omega\right)=\left(\Omega \exp\left[-2f\left(\Omega\right)\right]\right)^{n_f/4}$, $n$ is the break-up level and
$k$ is the polynomial order.

For a polynomial to approximate a matrix function it needs to approximate the
function well on the entire spectrum of the matrix. To determine these polynomials 
we employed a least square method \cite{Montvay1}. The coefficients $t_i$ of the polynomial
$T$ approximating the function $g$ are determined by minimizing the quadratic ``distance''
\begin{equation}
\delta ^{2}\left(t_{i}\right)=\int _{\lambda _{0}}^{\lambda _{1}}dx\, 
\rho (x)\left(T\left(x\right)-g\left(x\right)\right)^{2}, \label{distance}
\end{equation}
where $\lambda_0$, $\lambda_1$ are the spectral bounds of $\Omega$ and
$\rho$ is a weight function, ideally $\Omega$'s spectral density.

This method has the advantage that it turns into a linear problem and the solution is
\begin{equation}
t_{i}=\sum _{j=0}^{k}\left(K^{-1}\right)_{ij}g_{j},
\end{equation}
where $k$ is the polynomial order and
\begin{eqnarray}
K_{ij} & = & \int _{\lambda _{0}}^{\lambda _{1}}dx\, \rho \left(x\right)x^{i+j}, \nonumber \\
g_{i} & = & \int _{\lambda _{0}}^{\lambda _{1}}dx\, \rho \left(x\right)x^{i}g\left(x\right).
\end{eqnarray}

For our polynomials we used $\lambda_0 = 4m^2$, the lowest eigenvalue for the
staggered fermions matrix. We set the upper bound to $\lambda_1=16.4$ which we found to
be a good limit for the range of parameters we used in our simulations. It should be mentioned that
the only hard upper bound for the spectrum of $\Omega$ is $64+4m^2$ but this limit is reduced dynamically.
The smearing reduces the value of the highest eigenvalue even further. 

The weight function $\rho\left(x\right)$ should approximate the spectral
density for $\Omega$. However, we found that for reasonable weight functions the coefficients depend 
very little on the detailed form. Thus, we used a weight function of the form 
$\rho\left( x\right) = x^\alpha$ which is more convenient for our computation. The advantage of this form 
is that we can compute the $K_{ij}$ matrix analytically and also it generates a recursion relation for
the coefficients $g_i$.

\begin{figure}[t]
\includegraphics[height=0.9\columnwidth, width=1.7in, angle=-90]{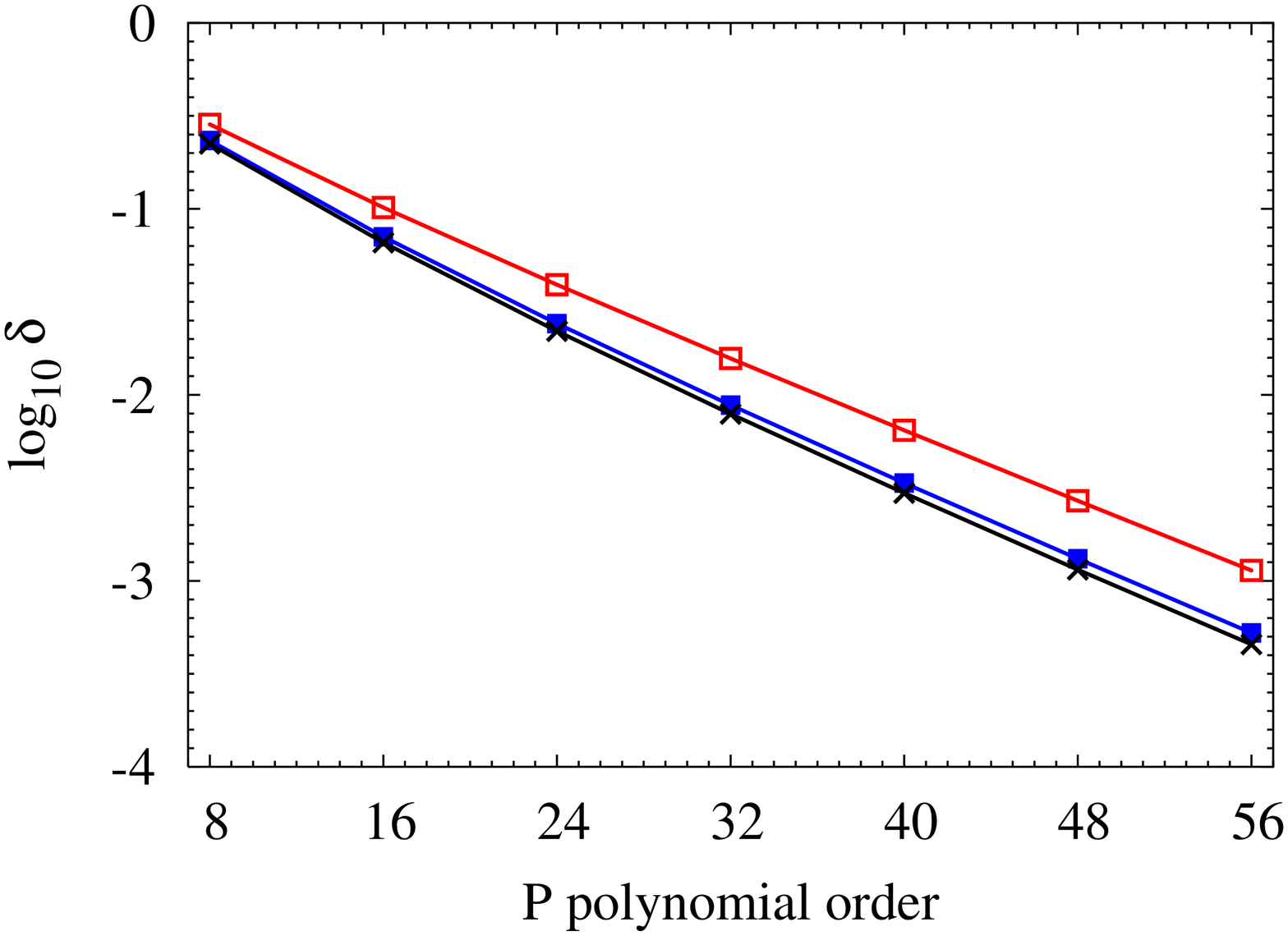}
\includegraphics[height=0.9\columnwidth, width=1.7in, angle=-90]{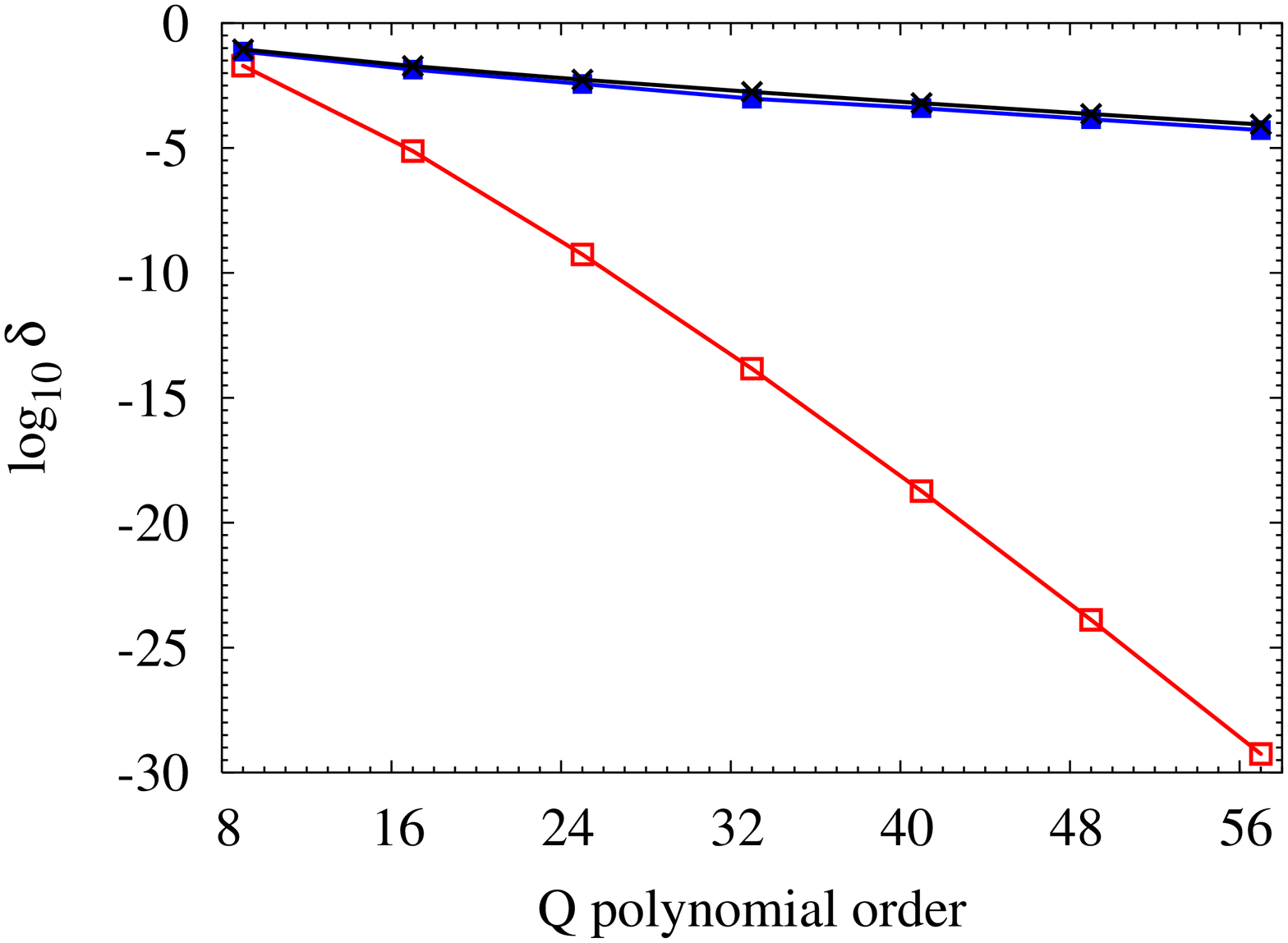}
\caption{The accuracy level for the P and Q polynomials:
open squares $n=1$, filled squares $n=4,$ crosses $n=8$ ($am=0.10$).}
\label{PQvsK}
\end{figure}

To measure the accuracy of the approximation we used the distance function (\ref{distance}) where we set
$\rho\left(x\right)=1$. In Fig.\ref{PQvsK} we plot the distance between 
$\left[P^{\left(n\right)}\right]^{2n}$, $\left[Q^{\left(n\right)}\right]^{n}$ and $F\left(x\right)^{\mp1}$ 
as a function of the polynomial order. We notice first that $Q^{\left(1\right)}$ is a much better
approximation than $Q^{\left(n\right)}$ for $n>1$. This is due to the fact that $F\left(x\right)$ 
(which is the function that $Q^{\left(1\right)}$ approximates) has an almost polynomial behavior
close to $\lambda_0$ where the function is the hardest to approximate. Another interesting feature that
we notice in the graphs is that the order of the approximation doesn't improve with increasing break-up level.
This is puzzling, at first, since as $n$ increases the equivalent order of the polynomial approximating 
$F\left(x\right)^{\pm 1}$ is $n\times k$. However, the number of variables varied to minimize $\delta$
stays the same and it seems that this is the determining factor for the order of approximation. 

We found that when $\delta\sim 10^{-8}$ the numerical roundoff error becomes dominant and the
higher order polynomials do not improve the precision of the calculation. Thus, we will be using the polynomials
with $\delta\sim 10^{-8}$ as ``exact'' ones. We see in Fig. \ref{PQvsK} that these polynomials have an order of $k\sim 100$.
The only exceptions are the polynomials for $n=1$ but they are of no real interest since they don't satisfy the 
minimum eigenvalue requirement.

\begin{figure}[t]
\includegraphics[height=.9\columnwidth, width=1.7in, angle=-90]{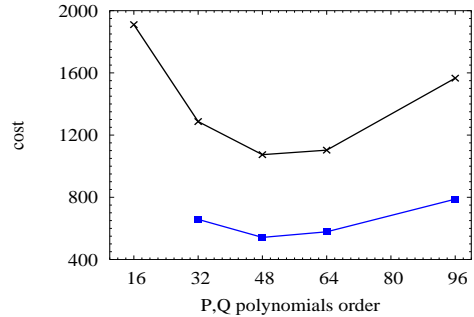}
\caption{Cost vs the polynomial order for $m=0.04$ and $n=4,8$.}
\label{cost}
\end{figure}

To further reduce the cost of computing the estimator we implemented a two step approach. Since we only use the 
value of the estimator in the accept reject step we only need to know whether the value of the estimator is larger
or smaller that $\ln r$, where $r$ is the random number. We will first compute the estimator approximatively, using
small order polynomials; if
the distance between the approximate value and $\ln r$ is greater than $\epsilon$ we use this value in the accept
reject step. If not, we ``fall-back'', i.e. we compute the ``exact'' value of the estimator using the large polynomials. 
The value of $\epsilon$ is determined in a small run as the maximum difference between the approximate estimator and 
its ``exact'' value. The rate of fall-back is usually around 10\% and this reduces the computational cost substantially.
Having fixed the large polynomials by the $\delta$ condition we determine the optimum small ones that minimize the
computational cost. The cost is given by the number of $M^\dagger M$ multiplies and it is determined by
the order of the polynomials used. We found that the order of the optimum small polynomial is independent of the 
break-up level (see Fig.\ref{cost}) and it increases as we decrease the mass. We find that 
for $am=0.10$ we need $k=24$, for $am=0.06$ $k=32$ and for $am=0.04$ $k=48$.

\section{CONCLUSIONS}

We have showed how to use the polynomial approximation for simulating dynamical fermions. First we determine
the minimum break-up level $n$ using the minimum eigenvalue requirement. Then we determine the large polynomials using
the $\delta\sim 10^{-8}$ conditions. We then determine the optimum small polynomials using the minimum cost condition.
For a more complete discussion see \cite{anna3}.

\end{document}